# Hidden structural and superconducting phase induced in antiperovskite arsenide SrPd$_3$As

*Akira Iyo,[1]\* Hiroshi Fujihisa,[1] Yoshito Gotoh,[1] Shigeyuki Ishida,[1] Hiroki Ninomiya,[1]*

*Yoshiyuki Yoshida,[1] Hiroshi Eisaki[1], Kenji Kawashima[1,2]*

[1]National Institute of Advanced Industrial Science and Technology (AIST), Tsukuba, Ibaraki

305-8568, Japan

[2] IMRA JAPAN Co., Ltd., Kariya, Aichi 448-8650, Japan

**ABSTRACT**

Enriching the material variation often contributes to the progress of materials science. We have discovered for the first time antiperovskite arsenide SrPd$_3$As and revealed a hidden structural and superconducting phase in Sr(Pd$_{1-x}$Pt$_x$)$_3$As. The Pd-rich samples ($0 \leq x \leq 0.2$) had the same non-centrosymmetric (NCS) tetragonal structure (a space group of $I4_1md$) as SrPd$_3$P. For the samples with $0.3 \leq x \leq 0.7$, a centrosymmetric (CS) tetragonal structure ($P4/nmm$) identical to that of SrPt$_3$P was found to appear, accompanied by superconductivity at a transition temperature ($T_c$) up to 3.7 K. In the samples synthesized with Pt-rich nominal compositions ($0.8 \leq x \leq 1.0$), Sr$_2$(Pd,Pt)$_{8-y}$As$_{1+y}$ with an intergrowth structure (CS-orthorhombic with $Cmcm$) was crystallized. The phase diagram obtained for Sr(Pd,Pt)$_3$As was analogous to that of (Ca,Sr)Pd$_3$P in that superconductivity ($T_c \geq 2$

K) occurred in the CS phases induced by substitutions to the NCS phases. This study indicates the potential for further material variation expansion and the importance of elemental substitutions to reveal hidden phases in related antiperovskites.

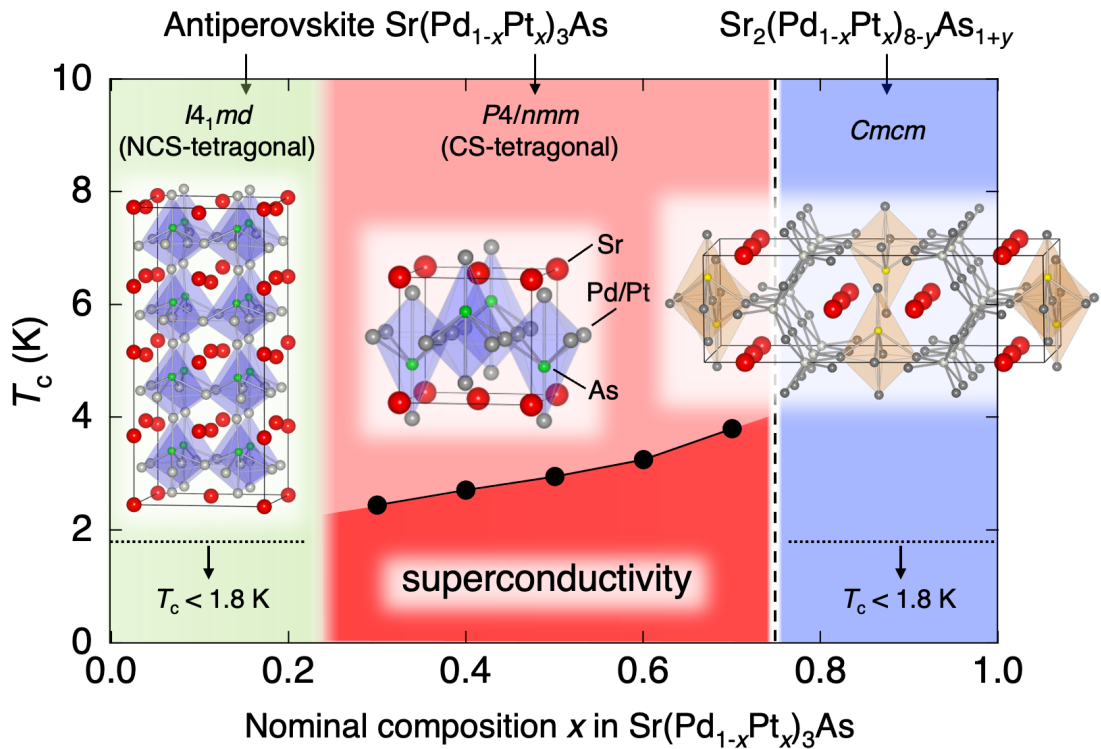

# 1. INTRODUCTION

Newly discovered superconducting materials often lead to the synthesis of new related materials, which can serve as the basis for advances in solid-state physics and chemistry and the development of applied materials. For example, the iron-based superconductor LaFePO with a transition temperature ($T_c$) of 4 K was discovered in 2006.[1] Then, in 2008, its As variant LaFeAsO system was reported to have a $T_c$ as high as 26 K.[2] This triggered a search for new related materials, and soon $T_c$ exceeded 50 K in the $Ln$FeAsO system, in which La was replaced by other lanthanides ($Ln$).[3] Subsequently, iron-based superconductors evolved into various material systems such as $R$Fe$_2$$Pn_2$ ($R$ = alkali and alkaline earth metals, $Pn$ = pnictogen) and Fe$Ch$ ($Ch$ = chalcogen).[3] Currently, the $R$Fe$_2$$Pn_2$- and Fe$Ch$-based superconductors are actively studied for applications in superconducting wires, bulks, and devices and as platform materials for exploring new physical phenomena.[4]

In 2019, an antiperovskite-related phosphide superconductor, Mg$_2$Rh$_3$P (a space group of $P4_132$), was reported.[5] Subsequently, related metal-rich phosphides were explored, and the antiperovskite $A$Pd$_3$P ($A$ = Ca, Sr, Ba) was discovered.[6,7] Although $A$Pd$_3$P did not exhibit superconductivity above 2 K, (Ca,Sr)Pd$_3$P, the solid solution between CaPd$_3$P and SrPd$_3$P, was found to have a structure ($Pnma$) different from that of CaPd$_3$P ($Aba2$) and SrPd$_3$P ($I4_1md$) and exhibit superconductivity with $T_c$ = 3.5 K. Furthermore, replacing divalent $A$ with trivalent La led to the synthesis of LaPd$_3$P ($I$-$43m$) with a new cubic prototype structure.[8] Interestingly, the space group ($P4/nmm$) observed in the known Pt variant $A$'Pt$_3$P ($A$' = Ca, Sr, La) [9] and the Pd-substituted Sr(Pt,Pd)$_3$P [10] did not appear in $A$Pd$_3$P, (Ca,Sr)Pd$_3$P, or LaPd$_3$P. Thus, the combination of elements and element substitutions in the antiperovskite phosphides gave rise to rich structural variations via the displacement of P and the deformation and rotation of octahedra in the structure.

Then, we were curious to determine whether the As variants of $A$Pd$_3$P and $A$'Pt$_3$P exist and have similar structural variations. As a result of expanding our material search to antiperovskite arsenides, a new antiperovskite arsenide SrPd$_3$As with a tetragonal structure was synthesized. Interestingly, the sample synthesized with the SrPt$_3$As composition had an orthorhombic structure, although the structure type was unknown. Based on the insight that a hidden phase can appear in a solid solution between antiperovskites of different space groups, which we learned from the study of $(Ca,Sr)Pd_3P$,[6] we attempted to synthesize a solid solution between $SrPd_3As$ and $SrPt_3As$ (neither material exhibited superconductivity above 1.8 K). Consequently, we found a hidden structural and superconducting phase in $Sr(Pd,Pt)_3As$. This study reports on the crystal structures and superconducting properties measured for $Sr(Pd_{1-x}Pt_x)_3As$ samples synthesized with various nominal compositions $x$.

## 2. Material and Methods

### 2.1 Material synthesis

Polycrystalline samples were synthesized via a solid-state reaction using starting materials of Pd powder (Kojundo Chemical, 99.9%), Pt powder (Kojundo Chemical, 99.9%), and a precursor with a nominal composition of SrAs. The SrAs precursor was obtained by reacting Sr (Furuuchi Chemical, 99.9%) with As (Kojundo Chemical, 99.9999%) at 750 °C for 24 h in an evacuated quartz tube. The Sr ingot was processed into flakes before the reaction to avoid unreacted Sr remaining in the precursor. The starting material was weighed to the nominal composition of $Sr(Pd_{1-x}Pt_x)_3As$ for $0 \leq x \leq 1$ in 0.1 increments and ground using a mortar in a glove box filled with $N_2$ gas. The ground powder was pressed into a pellet (~0.15 g) and enclosed in an evacuated quartz tube (inner diameter: 8 mm, length: ~70 mm). The sample was heated at 950 °C for $0 \leq x \leq 0.5$ and 920 °C for $0.6 \leq x \leq 1.0$, for a total reaction time of 6 h with an intermediate grinding followed by furnace cooling. The samples remained stable in the air.

### 2.2 Measurements

Powder XRD patterns were measured using a diffractometer (Rigaku, Ultima IV) with CuK$\alpha$ radiation (1.5418 Å) at approximately 293 K. The peak indexing was performed with the software BIOVIA Materials Studio (MS) X-Cell (version 2020 Revision 1).[11] The crystal structures were refined by Rietveld analysis using the MS Reflex software (version 2020 Revision 1).[12] Magnetization ($M$) was measured in a magnetic field ($H$) of 10 Oe using a magnetic property measurement system (Quantum Design, MPMS-XL7). The temperature ($T$) dependence of the electrical resistivity ($\rho$) was measured by a four-probe method in the range of 1.8–300 K using a physical property measurement system (Quantum Design, PPMS). The elemental composition of the samples was analyzed using an energy dispersive X-ray (EDX) spectrometer (Oxford, SwiftED3000) equipped with an electron microscope (Hitachi High-Technologies, TM-3000). Compositional analysis was performed on the flat polished surfaces of the sintered samples. The compositional ratios were determined by averaging the data obtained at more than 20 positions on the samples.

## 3. RESULTS AND DISCUSSION

### 3.1 Crystal structure analysis

Figure 1 shows the XRD patterns of the Sr(Pd$_{1-x}$Pt$_x$)$_3$As samples with the nominal composition of $x$ = 0, 0.2, 0.3, 0.5, 0.7, 0.8, and 1.0. Three distinct types of XRD patterns were observed depending on the composition $x$. Based on the structural analysis given below, the XRD patterns are assigned as tetragonal systems for $x$ = 0 and 0.2, $x$ = 0.3, 0.5, and 0.7, and an orthorhombic system for $x$ = 0.8 and 1.0.

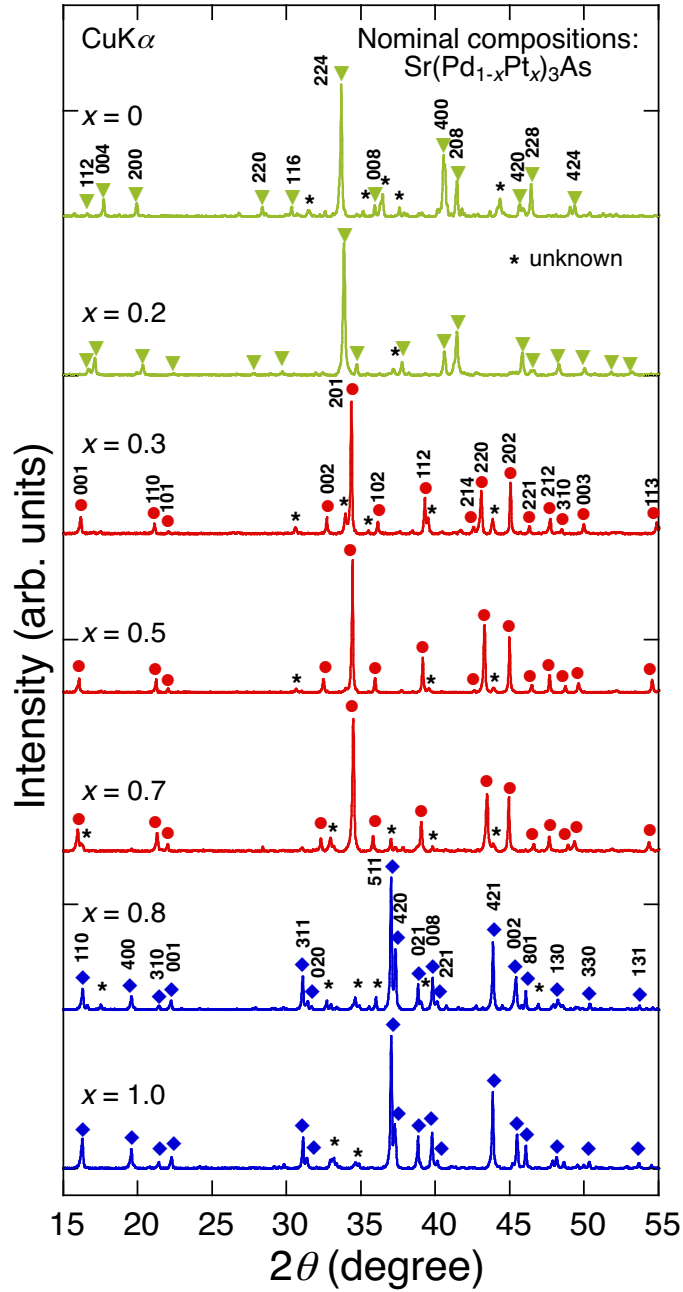

**Figure 1.** Powder XRD patterns of $Sr(Pd_{1-x}Pt_x)_3As$ samples synthesized with varying nominal composition $x$. Diffraction peaks are indexed as tetragonal systems for $x = 0$ and 0.2, $x = 0.3$, 0.5, and 0.7, and an orthorhombic system for $x = 0.8$ and 1.0. Unidentified peaks are indicated by asterisks.

The XRD patterns of the three representative samples synthesized with the nominal compositions $x = 0$, 0.6, and 1.0 were subjected to crystal structure analysis using the Rietveld method. The Rietveld fittings of the three XRD patterns are shown in Figures 2(a)–(c), and the refined structural parameters are summarized in Tables S1–3 of Supporting Information, respectively. The crystal structures determined by the Rietveld analysis are illustrated in Figures 3(a)–(c).

As shown in Figure 2(a), the XRD pattern of $SrPd_3As$ ($x = 0$) was well fitted with the same non-centrosymmetric tetragonal (NCSt) structural model ($I4_1md$) as that of $SrPd_3P$. The atoms of As enclosed in the $Pd_6As$ octahedra shift in the same direction along the $c$-axis, as shown in Figure 3(a), resulting in a lack of spatial inversion symmetry (non-centrosymmetry) in $SrPd_3As$. In addition, the rotation and deformation of the $Pd_6As$ octahedra caused the formation of a long-period structure with $Z$ (number of formula units in a unit cell) as large as 16.

Figure 2(b) shows that the XRD pattern of $Sr(Pd_{0.4}Pt_{0.6})_3As$ ($x = 0.6$) exhibited a good fit with the centrosymmetric tetragonal (CSt) structural model ($P4/nmm$), which is identical to $SrPt_3P$ ($BaAu_3Ge$-type). In this structure, the spatial inversion symmetry is preserved by displacing the neighboring As atoms in opposite directions along the $c$-axis (Figure 3(b)).

The elemental compositions of the samples synthesized with nominal compositions of $SrPd_3As$ and $Sr(Pd_{0.4}Pt_{0.6})_3As$ were analyzed to be $Sr_{0.98(2)}Pd_{3.00(4)}As_{1.00(4)}$ and $Sr_{1.00(3)}Pd_{1.07(5)}Pt_{1.93(3)}As_{1.03(3)}$ ($x = {\sim}0.64$), respectively. The analyzed values were close to the nominal compositions, and no evident crystallographic defects were detected. When site occupancies were included as fitting parameters for the refinements of $SrPd_3As$ and $Sr(Pd_{0.4}Pt_{0.6})_3As$, $R_{wp}$ decreased by only approximately 0.3% and 0.1%, respectively, and the atomic coordinates and occupancies did not change significantly. Therefore, the site occupancies were fixed at 1 for refinements.

For the Sr(Pd$_{0.4}$Pt$_{0.6}$)$_3$As sample, the error bar in the mixing ratio of Pd and Pt was estimated by the Rietveld analysis. The error in $x$ was analyzed as ±0.04, considering the range in which $R_{wp}$ produced a dominant error of 0.1%.

The sample synthesized with the nominal composition of SrPt$_3$As had a centrosymmetric orthorhombic (CSo) structure (*Cmcm*), as illustrated in Figure 3(c), which was analogous to the structure of Sr$_2$Pt$_{8-z}$As ($z$ = 0.715).[13] Sr$_2$Pt$_{8-z}$As has an incommensurately modulated structure due to the ordering of atomic vacancies in Pt sites. However, the assumption of the modulated structure for our sample deteriorated the fitting due to the appearance of extra peaks that were not observed experimentally. Therefore, we concluded that our sample did not have such a modulated structure.

The PtAs1 site is fully occupied by Pt or As in our model, and the chemical formula is described as Sr$_2$Pt$_{8-y}$As$_{1+y}$ or Sr$_2$(Pd$_{1-x}$Pt$_x$)$_{8-y}$As$_{1+y}$ for $x \neq 1$. Sr$_2$Pt$_{8-z}$As was synthesized under high pressure (2.3 GPa) with the same nominal composition as that of Sr$_2$Pt$_{8-y}$As$_{1+y}$. Therefore, the structural difference between Sr$_2$Pt$_{8-y}$As$_{1+y}$ and Sr$_2$Pt$_{8-z}$As can be attributed to the difference in the synthesis pressure.

The structural analysis of Sr$_2$Pt$_{8-y}$As$_{1+y}$ was performed by varying the mixing ratio of Pt and As in the PtAs1 site in 1% increments, and the fitting reliability factor was found to be minimum at 41(4)% Pt and 59(4)% As ($y$ = 1.18). The error bars for the mixing ratio $y$ were estimated by varying $y$ within a range where the increase in $R_{wp}$ was within 0.2%. Accordingly, the chemical composition of the sample obtained by the fitting was approximately Sr$_2$Pt$_{6.8}$As$_{2.2}$. An As-rich composition of Sr$_{2.12(4)}$Pt$_{6.80(4)}$As$_{2.08(4)}$ was obtained by the composition analysis, indicating the validity of the structural analysis. It should be noted that the mixing ratio must vary depending on the nominal composition and synthesis temperature.

Sr₂Pt₈₋ᵧAs₁₊ᵧ has an intergrowth structure by Sr(Pt,As)₅ (CaCu₅-type) and SrPt₃As (BaAu₃Ge-type) structural segments, as shown in Figure 3(d). The structure of Sr₂Pt₈₋ᵧAs₁₊ᵧ is closely related to that of Ce₂Pt₈P.[14] However, as indicated by the hatching of Pt₄As and Pt₄P pyramids in Figures 3(d) and (e), Ce₂Pt₈P includes the CePt₃P (CePt₃B-type) structural segment. Thus, the structure lacks spatial inversion symmetry (NCS orthorhombic (NCSo) with *Amm*2) in contrast to Sr₂Pt₈₋ᵧAs₁₊ᵧ.

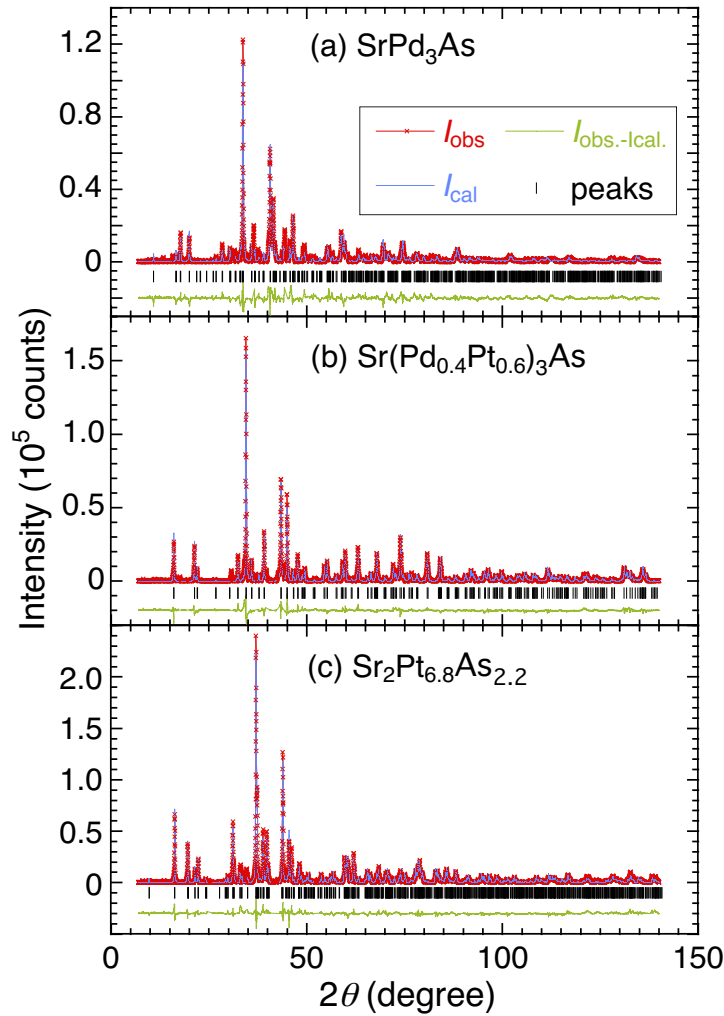

**Figure 2.** Rietveld fittings for (a) SrPd₃As, (b) Sr(Pd₀.₄Pt₀.₆)₃As, and (c) Sr₂Pt₆.₈As₂.₂ (Sr₂Pt₈₋ᵧAs₁₊ᵧ). $I_{obs.}$ and $I_{cal.}$ indicate the observed and calculated diffraction intensities, respectively. The fittings were performed on background-subtracted diffraction patterns.

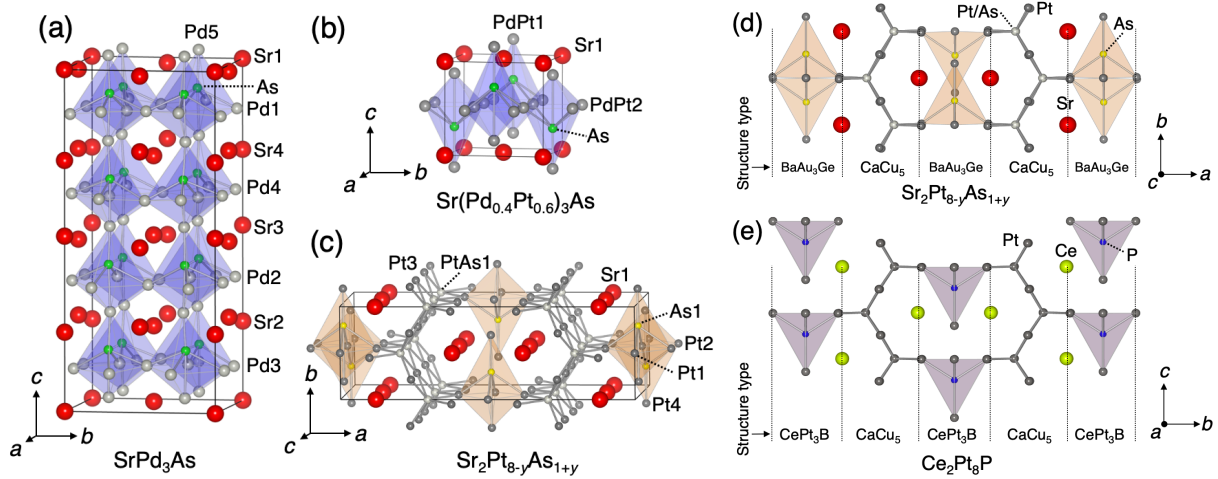

**Figure 3.** Crystal structures of (a) NCSt SrPd$_3$As, (b) CSt Sr(Pd$_{0.4}$Pt$_{0.6}$)$_3$As, and (c) CSo Sr$_2$Pt$_{8-y}$As$_{1+y}$. Unit cells are shown with solid lines, and octahedra enclosing As atoms are shown with hatching. (d) and (e) Crystal structures of CSo Sr$_2$Pt$_{8-y}$As$_{1+y}$ and NCSo Ce$_2$Pt$_8$P, respectively. It should be noted that the pyramids enclosing As(P) atoms are hatched for comparison. The crystal structure was drawn using the VESTA software.[15,17]

Figure 4 shows the reduced lattice parameters and cell volumes of the NCSt ($0 \leq x \leq 0.2$) and CSt ($0.3 \leq x \leq 0.7$) phases in Sr(Pd$_{1-x}$Pt$_x$)$_3$As together with those of CSo Sr$_2$(Pt$_{1-x}$Pd$_x$)$_{8-y}$As$_{1+y}$ ($0.8 \leq x \leq 1.0$). In each tetragonal phase, the $c$ parameter decreased while the $a$ parameter increased; the volume decreased with increasing Pt substitution. The same behavior of the lattice parameters and volume was reported for the tetragonal Sr(Pt,Pd)$_3$P.[10] The significant change in the structural parameters at the phase boundary between CSt and CSo was caused by the fundamental differences in their structure type and composition.

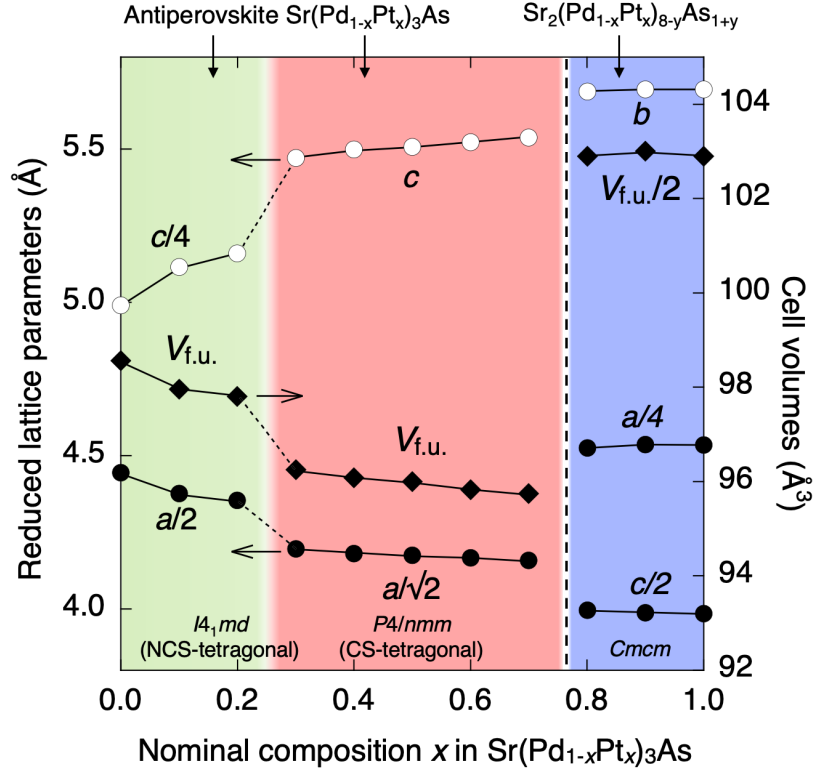

**Figure 4.** Reduced lattice parameters and cell volumes of compounds formed in the samples synthesized with the nominal composition of $Sr(Pd_{1-x}Pt_x)_3As$.

All three arsenides found in this study contained the $(Pd,Pt)_6As$ octahedra in the structures. Table 1 lists the structural parameters of the octahedra, together with the structurally corresponding phosphides for comparison.[6,9,14] Quadratic elongation $<\lambda>$ provides a quantitative measure of the distortion of octahedra, which is independent of the effective size.[17] The effective coordination number (ECoN) affords an approximation of the number of atoms coordinated to the central atom in polyhedra.[18] The definitions of $<\lambda>$ and ECoN have been provided in previous studies[17,18] and the manual of VESTA software.[17]

Several features can be derived from Table 1. The average bond length ($l_{av}$) and octahedral volume ($V_{oct}$) are larger for the arsenides than that for their corresponding phosphides because of the larger atomic size of As compared to that of P. In addition, the arsenides have larger $<\lambda>$ than their corresponding phosphides. The ECoN of As(P) was approximately 5, except for SrPd₃P. Therefore, it may be more appropriate to regard that the networks of pyramids, rather than octahedra, are formed in the structures, except for SrPd₃P. A comparison of the three arsenides revealed that the longer $l_{av}$ corresponds to the smaller $V_{oct}$. Although the result may appear contradictory, it is observed due to the increased $<\lambda>$.

**Table 1.** Structural parameters of the octahedra in the crystal structures for the three new arsenides and structurally related phosphides. VESTA software was used to derive the structural parameters.[15,16]

|  | SrPd₃As ($I4_1md$) | Sr(Pd₀.₄Pt₀.₆)₃As ($P4/nmm$) | Sr₂Pt₆.₈As₂.₂ ($Cmcm$) | SrPd₃P ($I4_1md$) | SrPt₃P ($P4/nmm$) | Ce₂Pt₈P ($Amm2$) |
|---|---|---|---|---|---|---|
| Average bond length ($l_{av}$) of As(P)-Pd(Pt) (Å) | 2.497 | 2.554 | 2.591 | 2.366 | 2.483 | 2.468 |
| Octahedral volume ($V_{oct}$) (Å³) | 16.22 | 15.96 | 15.30 | 15.46 | 15.14 | 14.57 |
| Quadratic elongation ($<\lambda>$) | 1.184 | 1.260 | 1.348 | 1.094 | 1.237 | 1.260 |
| Effective coordination number (ECoN) of As(P) | 4.652 | 5.035 | 4.738 | 5.722 | 4.909 | 4.635 |

### 3.2 Magnetization measurements

Figure 5(a) illustrates the $T$ dependence of $4\pi M/H$ below 4.5 K as a function of composition $x$. No diamagnetism due to superconductivity was detected for $0 \leq x \leq 0.2$. A superconducting transition was observed for $0.3 \leq x \leq 0.7$. As indicated by the arrows in Figure 5(a), $T_c$ increased with $x$, reaching a maximum value of 3.7 K at the phase boundary ($x = 0.7$). The samples with $x = 0.5$, 0.6, and 0.7 exhibited more than 90% of perfect diamagnetization ($4\pi M/H = -1$) at 2 K. However, the samples with $x = 0.3$ and 0.4 exhibited smaller diamagnetism. This result was possibly due to the poor sintering of the polycrystalline samples, which could not shield the applied magnetic field completely. In addition, the samples with $x = 0.3$ and 0.4 were more brittle than other samples. Superconductivity was suppressed below 1.8 K for $0.8 \leq x \leq 1.0$.

A phase diagram ($x$-dependent structure types and $T_c$ of the compounds formed in the samples) is shown in Figure 5(b). Only the CSt phase ($0.3 \leq x \leq 0.7$) showed superconductivity above 1.8 K. This phase diagram was analogous to that of (Ca,Sr)Pd$_3$P, as superconductivity ($T_c \geq 2$ K) appeared in the substitution-induced CS phase adjacent to the NCS phase.[6] We would like to stress that in (Ca,Sr)Pd$_3$P and Sr(Pd,Pt)$_3$As, the superconducting phase could not be found only by the synthesis of ternary endmembers. Two examples demonstrated the significance of element substitutions in inducing the hidden phases in the related antiperovskites.

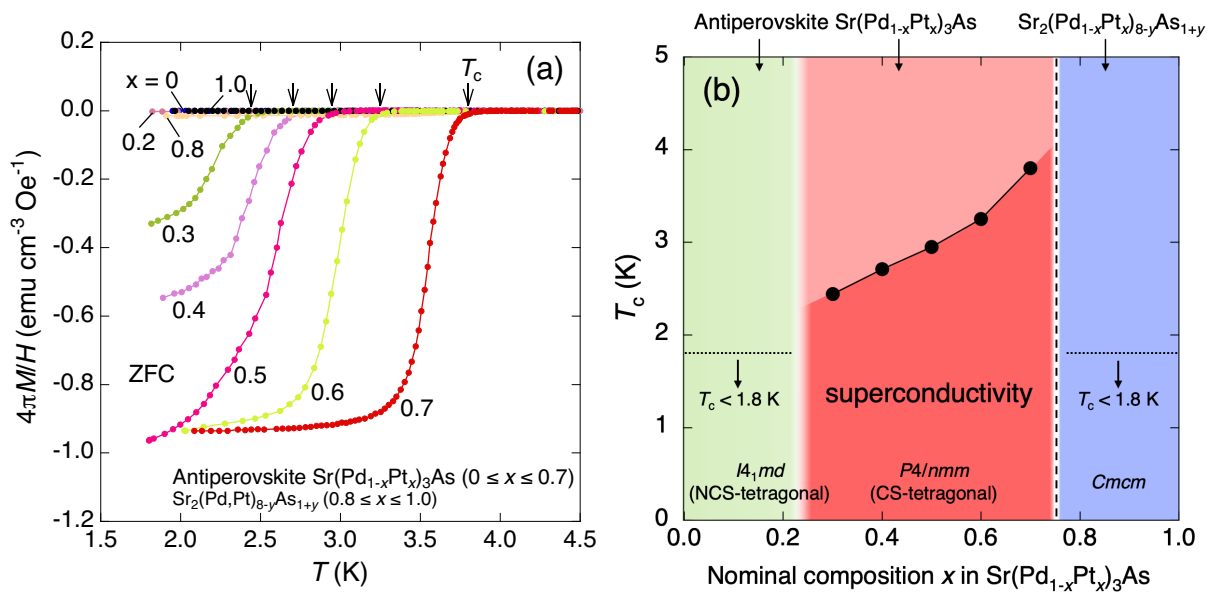

**Figure 5.** (a) $T$ dependence of $4\pi M/H$ for antiperovskite Sr(Pd$_{1-x}$Pt$_x$)$_3$As ($0 \leq x \leq 0.7$) and Sr$_2$(Pd$_{1-x}$Pt$_x$)$_{8-y}$As$_{1+y}$ ($0.8 \leq x \leq 1.0$). The measurements were performed in the zero-field cooling (ZFC) mode. Demagnetization corrections were applied to $M$ considering the sample shape. The arrows indicate the transition temperature ($T_c$). (b) The $T_c$ of the compounds formed in the samples synthesized with nominal compositions of Sr(Pd$_{1-x}$Pt$_x$)$_3$As.

### 3.3 Resistivity measurements

Figure 6(a) shows the $T$ dependence of $\rho$ normalized by the value at 300 K ($\rho^{300\,\text{K}}$) for samples with $0 \leq x \leq 1.0$. The $\rho^{300\,\text{K}}$ of each sample is shown in the inset. No anomaly was observed in the normal state $\rho$ for any sample, thus we assume that there was no structural phase transition in the measured temperature range.[6] As shown in the inset of Figure 6(a), the smallest $\rho^{300\text{K}}$ (0.56 m$\Omega$) of SrPd$_3$As ($x = 0$) might be due to the lack of disorder in crystal structure by the Pt substitution. The dome-shaped behavior of $\rho^{300\text{K}}$ in Sr(Pd$_{1-x}$Pt$_x$)$_3$As ($0 \leq x \leq 0.7$) were possibly due to the effect of disorder in crystal structure maximized at approximately $x = 0.5$.

As shown in Figure 6(b), clear superconducting transitions in resistivity were observed only for the CSt phase ($0.3 \leq x \leq 0.7$), with the highest onset $T_c$ of 4.0 K for $x = 0.7$. The sample with $x = 0.3$ did not exhibit zero resistance owing to the poor sintering, as suggested by the weak magnetic field shielding ability. The small drop in $\rho$ of approximately 4 K observed for $x = 0.8$ was due to the inevitable nonuniform Pd/Pt mixing in the polycrystalline sample.

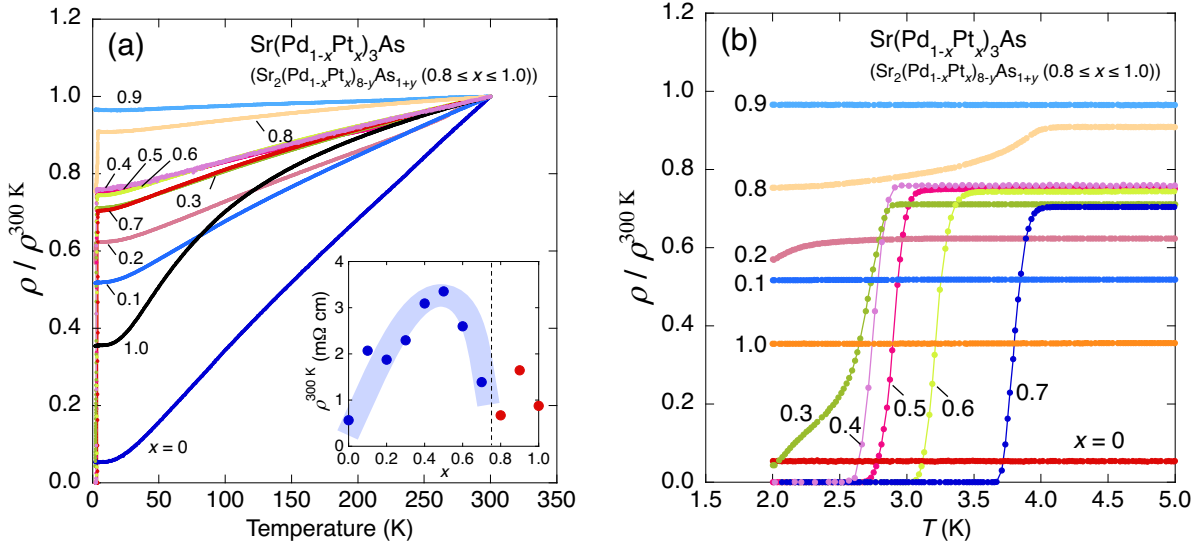

**Figure 6.** (a) *T* dependence of the normalized resistivity for Sr(Pd$_{1-x}$Pt$_x$)$_3$As ($0 \le x \le 0.7$) and Sr$_2$(Pd$_{1-x}$Pt$_x$)$_{8-y}$As$_{1+y}$ ($0.8 \le x \le 1.0$). The inset shows $\rho^{300\,\mathrm{K}}$ of each sample. (b) Magnified view near $T_c$.

## 4. CONCLUSIONS

We succeeded in synthesizing the antiperovskite arsenide SrPd$_3$As for the first time due to the attempt to expand the material variation of antiperovskites. In addition, based on the insights from the study of (Ca,Sr)Pd$_3$P, the hidden structural and superconducting phase was found in Sr(Pd,Pt)$_3$As. Superconductivity above 1.8 K ($T_c$ of approximately 3 K) occurred only in the substitution-induced CS phase as in the case of (Ca,Sr)Pd$_3$P. The discovery of antiperovskite arsenide is indicative of further new related materials to be explored. The antiperovskite phosphides/arsenides are of significant interest as they contain 4$d$ and/or 5$d$ transition metals, which are expected to yield unconventional superconductivity due to strong spin-orbit interactions. In addition, these antiperovskites have a unique feature that the inversion symmetry, a key for unconventional superconductivity, can be controlled through elemental substitutions. Thus, the antiperovskites can be promising materials for exploring new physical phenomena.

## AUTHOR INFORMATION

Corresponding Author

Akira Iyo (E-mail: iyo-akira@aist.go.jp)

Notes

The author declares no competing financial interest.

## ACKNOWLEDGMENT

This work was supported by the JSPS KAKENHI (No. JP19K04481 and JP19H05823).

**Supporting Information:** Refined structure parameters for $SrPd_3As$, $Sr(Pd_{0.4}Pt_{0.6})_3As$ and $Sr_2Pt_{6.8}As_{2.2}$ (PDF)